\documentclass[conference]{IEEEtran}
\IEEEoverridecommandlockouts
\usepackage{cite}
\usepackage{amsmath,amssymb,amsfonts}
\usepackage{algorithmic}
\usepackage[dvipdfmx]{graphicx}
\usepackage{textcomp}
\usepackage{xcolor}
\usepackage{bm}
\usepackage{url}
\usepackage{algorithm}
\def\BibTeX{{\rm B\kern-.05em{\sc i\kern-.025em b}\kern-.08em
    T\kern-.1667em\lower.7ex\hbox{E}\kern-.125emX}}

\begin{document}

\title{Physics-Aware Sparse Signal Recovery Through PDE-Governed Measurement Systems
}

\author{
\IEEEauthorblockN{Tadashi Wadayama}
\IEEEauthorblockA{ 
\textit{Nagoya Institute of Technology}\\
wadayama@nitech.ac.jp}
\and
\IEEEauthorblockN{Koji Igarashi}
\IEEEauthorblockA{ 
\textit{The University of Osaka}\\
iga.koji.es@osaka-u.ac.jp}
\and
\IEEEauthorblockN{Takumi Takahashi}
\IEEEauthorblockA{ 
\textit{The University of Osaka}\\
takahashi@comm.eng.osaka-u.ac.jp}

}

\maketitle

\begin{abstract}
This paper introduces a novel framework for physics-aware sparse signal recovery in measurement systems governed by partial differential equations (PDEs). Unlike conventional compressed sensing approaches that treat measurement systems as simple linear systems, our method explicitly incorporates the underlying physics through numerical PDE solvers and automatic differentiation (AD). We present physics-aware iterative shrinkage-thresholding algorithm (PA-ISTA), which combines the computational efficiency of ISTA with accurate physical modeling to achieve improved signal reconstruction. Using optical fiber channels as a concrete example, we demonstrate how the nonlinear Schrödinger equation (NLSE) can be integrated into the recovery process. Our approach leverages deep unfolding techniques for parameter optimization. Numerical experiments show that PA-ISTA significantly outperforms conventional recovery methods. While demonstrated on optical fiber systems, the proposed framework provides a general methodology for physics-aware signal recovery applicable to a wide range of various PDE-governed measurement systems.
\end{abstract}


\section{Introduction}

%
%
 
Over the past two decades, compressed sensing \cite{CS1,CS2} has revolutionized signal processing by enabling efficient reconstruction of signals from far fewer measurements than traditional sampling methods would require. This breakthrough has been particularly impactful in scenarios where signals exhibit sparsity, i.e., a property where the signal can be represented by only a few non-zero coefficients in an appropriate basis. 

The development of efficient sparse signal recovery algorithms \cite{Alg_survey} has been crucial to the practical success of compressed sensing. These algorithms typically solve an optimization problem that balances measurement fidelity with sparsity-promoting regularization. Notable examples include basis pursuit, least absolute shrinkage and selection operator (LASSO) \cite{LASSO,LARS}, and various iterative methods such as iterative shrinkage-thresholding algorithm (ISTA) \cite{ISTA,ISTA2}. These approaches have found applications in diverse fields, from medical imaging to astronomical observation.

However, many real-world inverse problems involve physical processes that cannot be adequately modeled by simple linear measurements. 
Examples include nonlinear optical effects like the Kerr effect in fiber optics, wave propagation in nonlinear acoustic media, and electromagnetic interactions with nonlinear meta-materials. 
In such cases, the reconstruction problem becomes significantly more challenging as it must account for the underlying physics governing signal propagation and measurement. 

There is growing recognition that incorporating explicit physical models into the inverse problem framework could potentially improve quality of solutions. 
	A prominent example is physics informed neural networks (PINNs) \cite{Raissi}. Beyond  solving forward problems, PINNs have proven particularly effective for partial differential equation (PDE)-based inverse problems, highlighting the significant potential of integrating physical models with machine learning methodologies.
This PINN approach represents a promising direction for applications where physical effects significantly influence the measurement process, although its full potential and limitations remain largely unexplored.

Recent advances in computational techniques, particularly automatic differentiation (AD) \cite{Baydin2018}, have opened new possibilities for incorporating physical models into signal processing frameworks. 
In this paper, we propose a general framework for {\em physics-aware sparse signal recovery} that incorporates PDEs describing the underlying physical phenomena. While our methodology is applicable to a wide range of physical observation systems governed by PDEs, we demonstrate its effectiveness using optical fiber channels \cite{Agrawal} as a concrete and challenging example.

\begin{figure}[htbp]
	\begin{center}
	\includegraphics[width=0.9\columnwidth]{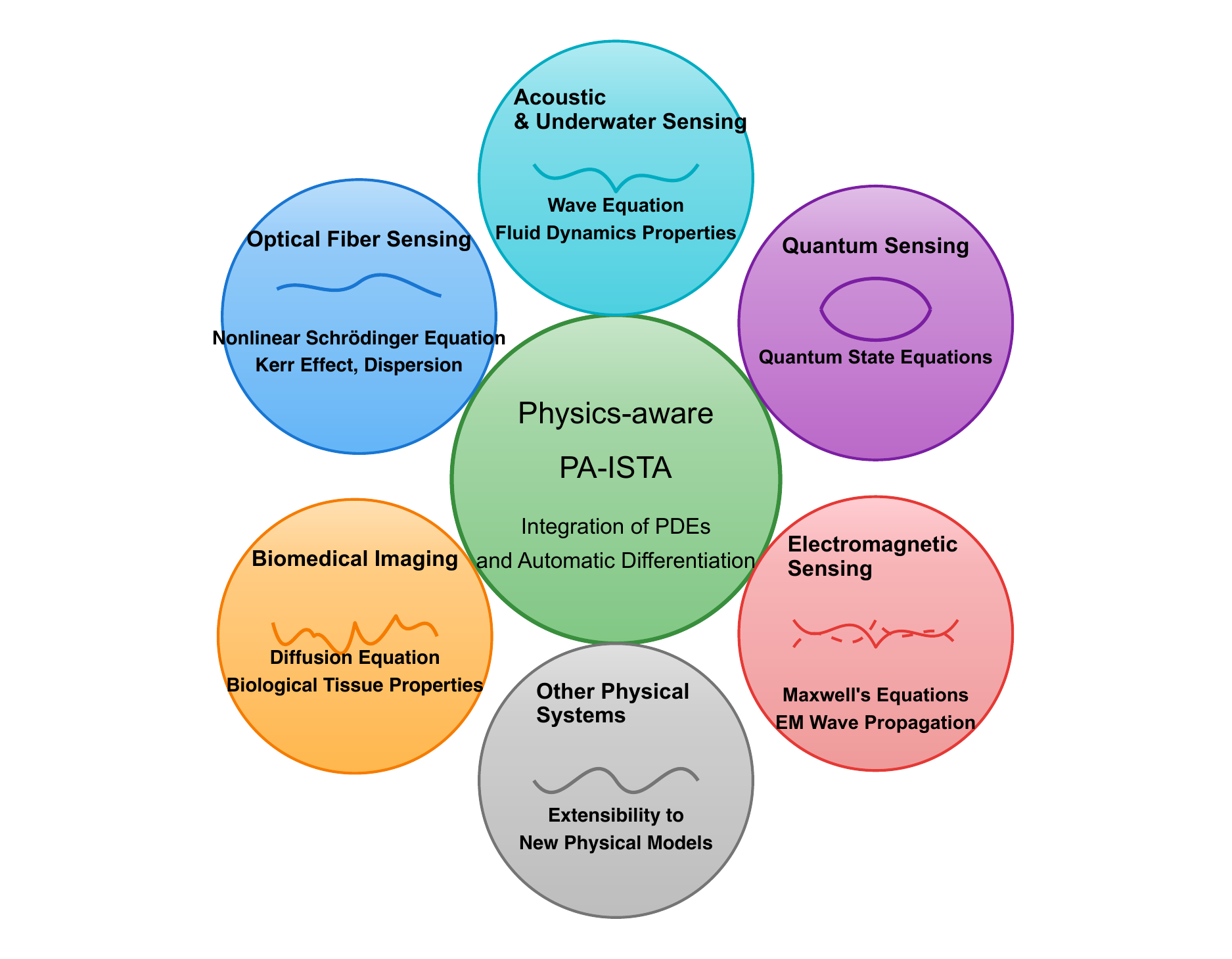}
	\caption{Scope for physics-aware sparse signal recovery.}
	\label{fig:overview}
	\end{center}
\end{figure}

In optical fiber systems, we integrate the computational efficiency of ISTA with numerical PDE solvers to achieve accurate signal reconstruction while faithfully capturing the underlying physics.
 The key innovation lies in our use of AD mechanism to compute gradients through a numerical PDE solver, 
 thereby facilitating efficient optimization despite the complexity of the physical system. 
 The proposed framework is general enough to be adapted to other physical systems 
 where the underlying physics can be described by PDEs, such as heat conduction processes, 
 wave propagation in elastic media, Maxwell's equations for electromagnetic waves, or fluid dynamics.
 Figure \ref{fig:overview} shows the scope for physics-aware sparse signal recovery presented in this paper.

As a related work, the concept of {\em physics-aware decoding} that successfully uses automatic 
differentiation in a decoding process for binary linear codes has been presented in \cite{ISIT2025}. 
However, the extension to
sparse signal recovery problems has not yet been discussed so far.
The main contributions of this paper are as follows:
\begin{itemize}
	\item We propose a general framework for physics-aware sparse signal recovery that incorporates PDEs describing the underlying physical phenomena.
	\item We demonstrate the effectiveness of the proposed framework using optical fiber channels as a concrete example.
	\item Numerical experiments show that the proposed framework significantly outperforms conventional naive recovery methods.
\end{itemize}

The rest of this paper is organized as follows. 
In Section \ref{sec:preliminaries}, we review the fundamentals of compressed sensing and the ISTA algorithm. 
In Section \ref{sec:signal_recovery}, we formulate the sparse signal recovery problem and introduce the proposed framework. 
In Section \ref{sec:experiments}, we present numerical experiments to demonstrate the effectiveness of the proposed framework. 
In Section \ref{sec:conclusion}, we draw conclusions and discuss future work.

We have released an implementation, including scripts to reproduce numerical results 
for sparse signal recovery, at \url{https://github.com/wadayama/PA-ISTA}.

\section{Preliminaries}
\label{sec:preliminaries}
\subsection{Sparse signal recovery problem with linear measurement}

In this subsection, we review the fundamentals of compressed sensing with simple linear measurements. Let us consider a sparse signal $\bm{s} \in \mathbb{R}^n$, where sparsity implies that only a small fraction of its elements are non-zero relative to the dimension $n$. The measurement process can be described by a linear model:
\begin{align}
\bm{y} = \bm{A} \bm{s} + \bm{w},	
\end{align}
where $\bm{y} \in \mathbb{R}^m$ $(m < n)$ represents the measurement vector, $\bm{A} \in \mathbb{R}^{m \times n}$ is the measurement matrix, and $\bm{w} \in \mathbb{R}^m$ denotes the measurement noise.
In the sparse signal recovery problem, the objective is to recover the original signal $\bm{s}$ 
from the given measurement vector $\bm{y}$ with maximum possible accuracy. The observer has access only to $\bm{y}$ and $\bm{A}$. 
It is important to note that since $m < n$, without considering the sparsity of the original signal $\bm{s}$, 
this becomes an underdetermined problem with no unique solution.
This framework is widely applicable such as MRI for medical imaging, and grant-free wireless communication systems.

A widely adopted approach to sparse signal reconstruction is the Lasso  \cite{LASSO,LARS}, which formulates the problem as a regularized least squares optimization. The Lasso framework reconstructs the sparse signal by solving the convex optimization problem:
\begin{align}
\bm{\hat s} \equiv \mbox{argmin}_{\bm{x}\in\mathbb{R}^{n}} \left( \frac{1}{2} \|\bm{y} - \bm{A}\bm{x}\|_2^2 + \lambda \|\bm{x}\|_1 \right),	
\end{align}
where $\bm{\hat s}$ represents the reconstructed signal and $\lambda (> 0)$ is a regularization parameter. The objective function consists of two terms: a quadratic data fidelity term that measures the reconstruction error, and an L1-norm regularization term that promotes sparsity in the solution. The parameter $\lambda$ controls the trade-off between these competing objectives.

\subsection{Related works}
Various discrete-time algorithms for sparse signal 
recovery~\cite{Alg_survey}
 have been proposed 
in the literature based on the Lasso framework~\cite{LASSO, MP, LARS}.
Among them, the iterative shrinkage thresholding algorithm (ISTA)~\cite{ISTA, ISTA2} 
has emerged as one of the most widely used approaches for solving the Lasso problem.
ISTA operates by iteratively applying two key processes:
a linear estimation step followed by a shrinkage operation using a soft thresholding function.
From an optimization perspective, ISTA can be interpreted as a proximal gradient descent method~\cite{Prox}
that naturally arises from the Lasso formulation.
The approximate message passing (AMP) algorithm~\cite{reBP,AMP} represents another significant advance,
offering substantially faster convergence compared to ISTA.
A notable extension is the orthogonal AMP (OAMP)~\cite{OAMP} developed by Ma and Ping,
which broadens the applicability to diverse sensing matrix classes,
particularly those exhibiting unitary invariance.
Further theoretical progress within the Bayesian regime was achieved by Rangan et al. with vector AMP (VAMP)~\cite{VAMP},
provides rigorous theoretical guarantees through state evolution analysis.

Signal detection algorithms play a crucial role in receiver design for optical fiber communication systems.
Digital back propagation (DBP) \cite{Ip} is a well-established 
signal detection technique that leverages the NLSE solver in reverse way. 
The learnable variants of DBP introduced in \cite{Haeger1, Haeger2} 
represent pioneering contributions demonstrating the applicability of deep learning 
to signal processing in optical fiber communications.

\subsection{ISTA for sparse signal recovery}

ISTA \cite{ISTA,ISTA2}
is a well-known proximal gradient method for solving 
the Lasso problem. In this section, we aim to derive the ISTA update equations.
For the Lasso minimization problem, let us define
$
f(\bm{x}) \equiv \frac{1}{2} \| \bm{y} - \bm{A} \bm{x} \|_2^2,
$
and
$
h(\bm{x}) \equiv \lambda\|\bm{x}\|_1.
$
It is important to note that since the L1 regularization term is non-differentiable, algorithms that assume differentiability of the objective function cannot be applied.
The gradient vector of the squared error term $f(\bm{x})$ is given by
\begin{align}
   \nabla f(\bm{x}) = \bm{A}^T (\bm{A} \bm{x} - \bm{y}).	
\end{align}
The proximal operator \cite{Prox} of the L1 regularization term $\tau \|\bm{x}\|_1$,
\begin{align}
    \mbox{prox}_{\tau \|\bm{x}\|_1}(\bm{x}) = S_{\tau}(\bm{x})
    \equiv \mbox{sign}(x) \max \{|x| - \tau, 0 \},
\end{align}
is the soft-thresholding function. 
In sparse signal reconstruction problems, the proximal operator corresponding to the regularization term is sometimes called the {\em shrinkage function}.

ISTA is a proximal gradient method \cite{Prox} 
derived from the Lasso problem and is defined by the iterative equations:
\begin{align}
    \bm{z}^{(k)} &= \bm{x}^{(k)} - \eta \bm{A}^T (\bm{A} \bm{x}^{(k)} - \bm{y}) \label{gradient_ista} \\
    \bm{x}^{(k+1)} &= S_{\eta \lambda}(\bm{z}^{(k)}), \quad k=0,1,\dots \label{soft_ista}
\end{align}
Eq.~(\ref{gradient_ista}) is called the {\em gradient descent step}, 
and Eq.~(\ref{soft_ista}) is called the {\em shrinkage step}.
Here, the step size parameter $\eta$ is a real constant, and ISTA converges to the Lasso solution when
$
    \eta < 1/\Gamma
$
where $\Gamma$ is the maximum eigenvalue of the Gram matrix $\bm{A}^T \bm{A}$.

\subsection{Nonlinear Schrödinger Equation (NLSE)}
We consider the NLSE:
\begin{align} \label{NLSE}
	\frac{\partial U}{\partial z}  = - \frac{i \beta_2}{2} \frac{\partial^2 U}{\partial t^2} + i \gamma |U|^2 U,  
\end{align}
where $i$ denotes the imaginary unit \cite{Agrawal}. The variables $z$ and $t$ represent 
the position in a fiber and time, respectively.
The function $U(t,z)$ describes the optical field in the optical fiber.
The NLSE plays a fundamental role in single-mode optical fiber communications, where it governs signal propagation by describing the evolution of optical pulse shapes and phases along the fiber length. This evolution is influenced by physical effects such as dispersion and nonlinearity. The parameter $\beta_2 \in \mathbb{R}$ is the dispersion constant
	and the nonlinear coefficient $\gamma  \in \mathbb{R}$ characterizes the strength of the fiber nonlinearity.

\subsection{Split-State Fourier Method (SSFM)}

The NLSE involves both linear dispersion terms and nonlinear effects, making it challenging to solve directly. The split-step Fourier method (SSFM) \cite{Agrawal} provides an efficient numerical approach by leveraging the distinct characteristics of these terms. Consider the NLSE:
\begin{align}
    {\partial A}/{\partial z} = (\hat{D} + \hat{N})A,
\end{align}
where $A$ is the complex envelope of the optical field, $\hat{D}$ represents the linear dispersion operator in the frequency domain, and $\hat{N}$ represents the nonlinear operator in the time domain.

The fundamental principle of SSFM lies in its approach to handling operators that cannot be solved simultaneously. While the combined effect of dispersion and nonlinearity cannot be computed exactly, SSFM provides an efficient approximation by treating these effects separately over small propagation steps $\Delta z$. The method iterates the following stages at each step:
\begin{enumerate}
	\item The signal is transformed to the frequency domain using Fast Fourier Transform (FFT);
	\item The dispersion effect is applied in the frequency domain using the operator $\hat{D}$;
	\item The signal is then transformed back to the time domain through inverse FFT;
	\item The nonlinear effects are computed in the time domain using the operator $\hat{N}$;
	\item This process repeats for the next propagation step.
\end{enumerate}

This approach is particularly efficient because the dispersion operator $\hat{D}$ takes a simple multiplicative form in the frequency domain, while the nonlinear operator $\hat{N}$ is easily computed in the time domain. The FFT provides an efficient means to switch between these domains, making the overall method computationally practical.
The symmetrized version of SSFM, which applies half of the linear operation before and after the nonlinear step, achieves second-order accuracy with respect to the step size $\Delta z$. Specifically, it approximates the solution over a step $\Delta z$ as
\begin{align}
    A(z+\Delta z) \approx \exp\left(\frac{\Delta z}{2}\hat{D}\right)\exp \left (\Delta z\hat{N} \right)\exp \left(\frac{\Delta z}{2}\hat{D} \right)A(z).
\end{align}
This improved accuracy, combined with its computational efficiency, makes the symmetrized SSFM particularly suitable for our physics-aware recovery framework where both precision and computational tractability are essential.
A pseudo-code of the SSFM is given in Algorithm \ref{alg:ssfm} according to \cite{Agrawal}.

\begin{algorithm}
	\caption{Split-Step Fourier Method (SSFM) for NLSE}
	\label{alg:ssfm}
	\begin{algorithmic}[1]
	\REQUIRE Input waveform $U(t, 0)$, fiber length $L$, spatial step size $\Delta z$, temporal grid $\{t_1, t_2, \ldots, t_{N_t}\}$, parameters $\beta_2$, $\gamma$
	\ENSURE Output waveform $U(t, L)$
	\STATE $N_z \gets \lfloor L / \Delta z \rfloor$ 
	\STATE $U_{\text{current}} \gets U(t, 0)$ 
	\FOR{$k = 1$ to $N_z$}
		\STATE $z \gets k \cdot \Delta z$ 
		
		\STATE // Linear step (frequency domain) - first half
		\STATE $\tilde{U} \gets \text{FFT}(U_{\text{current}})$ 
		\FOR{$i = 1$ to $N_t$}
			\STATE $\omega_i \gets 2\pi (i - N_t/2) / (N_t \cdot \Delta t)$ 
			\STATE $\tilde{U}_i \gets \tilde{U}_i \cdot \exp\left(\frac{i\beta_2\omega_i^2\Delta z}{4}\right)$ 
		\ENDFOR
		\STATE $U_{\text{mid}} \gets \text{IFFT}(\tilde{U})$ 
		
		\STATE // Nonlinear step (time domain)
		\FOR{$i = 1$ to $N_t$}
			\STATE $U_{\text{mid},i} \gets U_{\text{mid},i} \cdot \exp(i\gamma|U_{\text{mid},i}|^2\Delta z)$ 
		\ENDFOR
		
		\STATE // Linear step (frequency domain) - second half
		\STATE $\tilde{U} \gets \text{FFT}(U_{\text{mid}})$ 
		\FOR{$i = 1$ to $N_t$}
			\STATE $\omega_i \gets 2\pi (i - N_t/2) / (N_t \cdot \Delta t)$ 
			\STATE $\tilde{U}_i \gets \tilde{U}_i \cdot \exp\left(\frac{i\beta_2\omega_i^2\Delta z}{4}\right)$ 
		\ENDFOR
		\STATE $U_{\text{current}} \gets \text{IFFT}(\tilde{U})$ 
	\ENDFOR
	\STATE \RETURN $U_{\text{current}}$ 
	\end{algorithmic}
\end{algorithm}

\section{Sparse Signal Recovery Problem for PDE-governed Measurement Systems}
\label{sec:signal_recovery}
\subsection{Overview}

Although our physics-aware sparse signal recovery framework is applicable to a wide range of partial differential equations, we demonstrate its effectiveness using the NLSE (\ref{NLSE}) as a concrete example. In our problem formulation, we consider a system where a sensing device at the fiber input ($z = 0$) injects a waveform $U(t, 0)$ consisting of several sparse pulses, which can be considered as the {\em target sparse signals}. As this signal propagates through the fiber, it is observed by a detector placed at the fiber output ($z = L$). Our goal is to reconstruct the target sparse signal from these output measurements, effectively recovering the input signal through the nonlinear fiber channel.
Specifically, as the optical signal propagates through the fiber, the waveform evolves according to the NLSE (\ref{NLSE}) with increasing $z$.
The signal recovery problem can be stated as follows: given a noisy observation of the waveform $U(t, L)$ at position $z = L$, estimate the original sparse signal $U(t, 0)$. In other words, we aim to detect the set of sparse pulses at the fiber input ($z = 0$) using measurements obtained at the fiber output ($z = L$).

\subsection{Details}

\begin{figure}
\begin{center}
\includegraphics[width=\columnwidth]{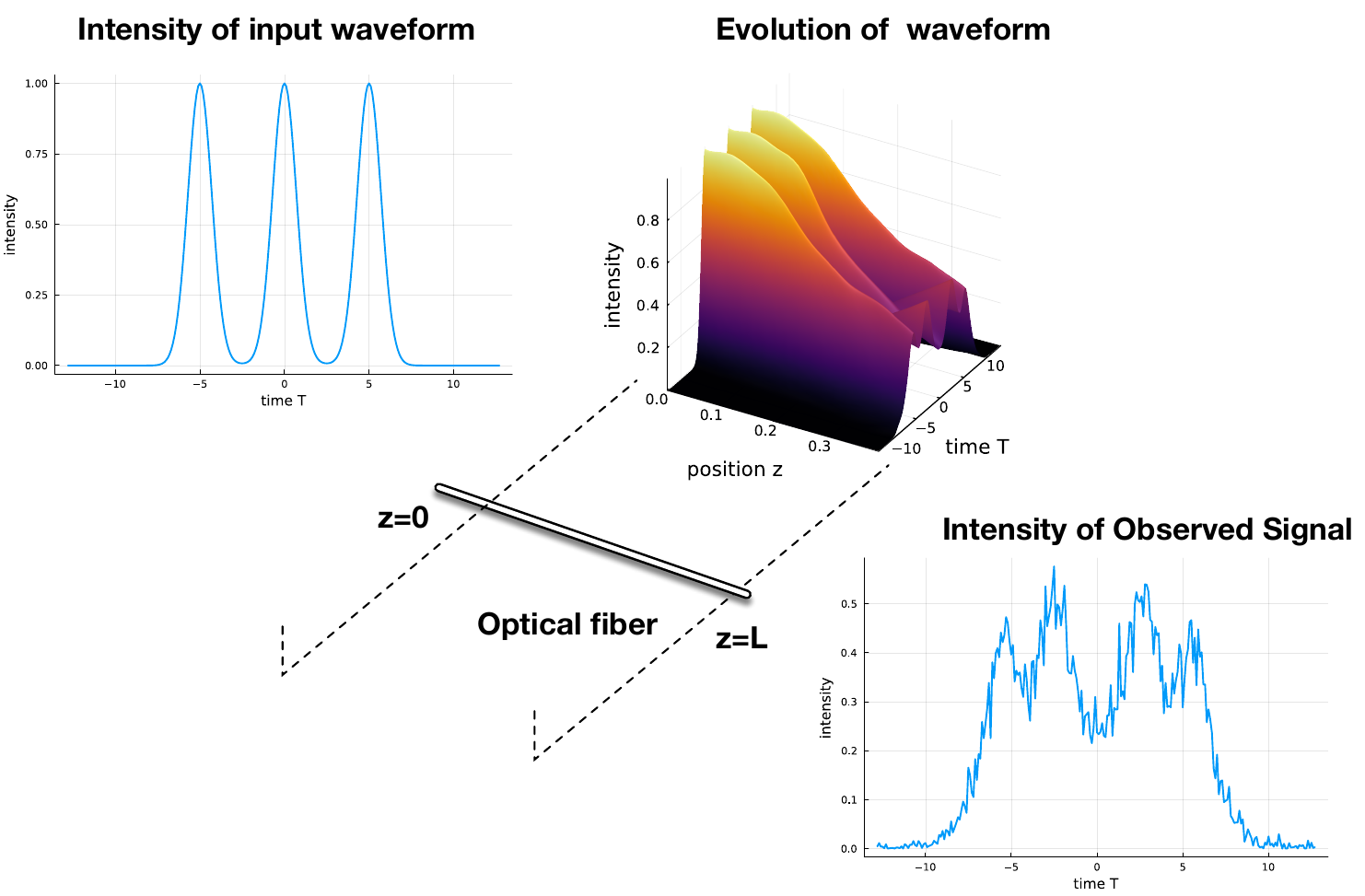}
\caption{Visualization of signal propagation through an optical fiber. The left panel displays the intensity profile of the input signal $|U(t,0)|^2$ at the fiber input. The center panel shows a three-dimensional visualization of the signal intensity $|U(t,z)|^2$, illustrating how the waveform evolves as it propagates along the fiber length $z$. The right panel presents the measured intensity profile $|U(t,L)|^2$ at the fiber output, including the effects of measurement noise.}
\label{fig:fiber_model}
\end{center}
\end{figure}

Let us define a Gaussian-shaped pulse function as
\begin{align}
\phi(x) \equiv \exp\left(-\frac{x^2}{2 T_0^2} \right),			
\end{align}
where $T_0$ represents the pulse half-width (measured at the $1/e$-intensity point). We model the initial waveform as a linear combination of Gaussian-shaped pulses $\phi(x-p_i)$ centered at positions ${p_i } (i \in [n])$. Specifically, the input waveform at the fiber input ($z = 0$) is given by
\begin{align}
U(\tau, 0) =  \sum_{i=1}^n s_i \phi(\tau - p_i),		
\end{align}
where $\bm s \equiv (s_1,s_2,\ldots,s_n) \in \mathbb{C}^n$ represents a sparse complex vector with only $k$ non-zero components ($k \ll n$). This input waveform serves as the boundary condition for the NLSE (\ref{NLSE}). 
Let $U(t, z; \bm s)$ denote the unique solution of the NLSE (\ref{NLSE}) that satisfies this boundary condition.
At the fiber output, the detector performs 
measurements by sampling the waveform 
at specific time points $q_i (i \in [m])$, yielding the sampled values:
\begin{align}
y_i = U(q_i, L;\bm s) + n_i, \quad i \in [m],	
\end{align}
where each noise term $n_i$ follows a complex Gaussian distribution, i.e., 
$n_i \sim {\cal CN}(\bm 0, \sigma^2)$ where $\sigma^2$ denotes noise variance.
Figure \ref{fig:fiber_model} visualizes 
behavior of signal propagation through an optical fiber.

Based on the above formulation, we can now precisely define our sparse signal recovery problem. Given the observation vector $\bm y \equiv (y_1,y_2,\ldots, y_m) \in \mathbb{C}^m$ at the fiber output, our objective is to accurately reconstruct the original sparse complex vector $\bm s$. This recovery problem presents significant challenges due to the nonlinear nature of the wave evolution governed by the NLSE, making it fundamentally different from conventional sparse recovery problems.

\section{Proposed Method}
\label{sec:proposed}

\subsection{Lasso-like formulation}

To address the sparse signal recovery problem described above, it is natural to consider an optimization-based approach analogous to the classical Lasso formulation. We propose to minimize a Lasso-like objective function $F:\mathbb{C}^n \rightarrow \mathbb{R}$ defined as
\begin{align}
F(\bm s) \equiv \|\bm y - f(\bm s) \|^2_2 + \lambda \|\bm s\|_1,			
\end{align}
where 
\begin{align}
f(\bm s) \equiv (U(q_1, L;\bm s),\ldots, U(q_m, L;\bm s)) 
\end{align}
represents the noiseless output of the nonlinear fiber channel at the sampling points. The first term measures the fidelity between the observed samples and the predicted output, while the second term promotes sparsity in the solution. By minimizing this objective function, we aim to obtain a sparse solution that is consistent with both the observed data and the underlying physics described by the NLSE.

\subsection{Physics-aware ISTA}
The optimization problem formulated in the previous subsection can be expressed as
\begin{align}
\mbox{minimize}_{\bm s \in \mathbb{C}^n} F(\bm s),	
\end{align}
where the objective function becomes non-convex when the optical fiber channel exhibits significant nonlinearity. Despite this non-convexity, we propose to adapt the ISTA framework to minimize $F(\bm s)$. This leads to a complex-valued variant of the ISTA algorithm:
\begin{align} \label{complex_gd_step}
\bm{z}^{(k)} &= \bm{x}^{(k)} - \eta^{(k)} \nabla_{\bm{x}^{(k)}} \|\bm y - f(\bm{x}^{(k)}) \|^2_2 \\
\bm{x}^{(k+1)} &= T_{\theta^{(k)}}(\bm{z}^{(k)}), \quad k=0,1,\dots, \label{ista_like}
\end{align}
where $T_\tau: \mathbb{C} \rightarrow \mathbb{C}$ denotes the complex shrinkage operator defined by
\begin{align}
T_\tau(x) \equiv \frac{x}{|x|} \{\max |x| - \tau, 0 \}.		
\end{align}

It is important to note that the objective function $F$ defined over the complex field $\mathbb{C}$ is non-holomorphic. Consequently, when computing the gradient
$\nabla_{\bm{x}^{(k)}} \|\bm y - f(\bm{x}^{(k)}) \|^2_2$,
we must employ the {\em Wirtinger derivative}, which provides the appropriate framework for differentiation of non-holomorphic functions.
Wirtinger calculus provides a rigorous framework for optimization in the complex domain. This approach treats a complex variable $z$ and its conjugate $z^*$ as independent variables, defining the derivatives:
\begin{align}
    \frac{\partial g}{\partial z} = \frac{1}{2}\left(\frac{\partial g}{\partial x} - i\frac{\partial g}{\partial y}\right),\ 
    \frac{\partial g}{\partial z^*} = \frac{1}{2}\left(\frac{\partial g}{\partial x} + i\frac{\partial g}{\partial y}\right)
\end{align}
for a real-valued function $g(z)$
where $z = x + iy$. These Wirtinger derivatives enable gradient-based optimization of our non-holomorphic objective function in the complex domain.
Namely, the conjugate derivative can be used in gradient descent processes.
In our implementation, AD naturally computes these Wirtinger derivatives when operating on complex variables.

A significant implementation challenge lies in computing the gradient term in (\ref{complex_gd_step}), since the NLSE (\ref{NLSE}) rarely admits closed-form solutions. To address this challenge, we employ 
a numerical PDE solver, i.e., the SSFM solver,  to approximate the channel output $f$. The approximate noiseless output at the sampling points is defined as
\begin{align}
\hat{f}(\bm s) \equiv (\hat{U}(q_1, L;\bm s), \ldots, \hat{U}(q_m, L;\bm s)),	
\end{align}
where $\hat{U}(\cdot,\cdot ;\bm s)$ represents the approximate solution obtained through the SSFM solver. By leveraging AD, we can efficiently compute the Wirtinger gradient 
$\nabla_{\bm{x}^{(k)}} \|\bm y - \hat{f}(\bm{x}^{(k)}) \|^2_2$. This leads to a recursive formula of {\em physics-aware ISTA (PA-ISTA) algorithm}:
\begin{align} 
\bm{z}^{(k)} &= \bm{x}^{(k)} - \eta^{(k)} \nabla_{\bm{x}^{(k)}} \|\bm y - \hat{f}(\bm{x}^{(k)}) \|^2_2 \\
\bm{x}^{(k+1)} &= T_{\theta^{(k)}}(\bm{z}^{(k)}), \quad k=0,1,\dots. \label{PA-ISTA}
\end{align}

The complete description of physics-aware sparse signal recovery algorithm is given 
in Algorithm \ref{alg:pa-ista}. The choice of the squared error function as 
a loss function is motivated by our assumption of additive white Gaussian noise in the system model. 
For non-Gaussian noise scenarios, the error function can be appropriately modified to match the underlying noise statistics.

\begin{figure}[htbp]
	\begin{center}
	\includegraphics[width=1.0\columnwidth]{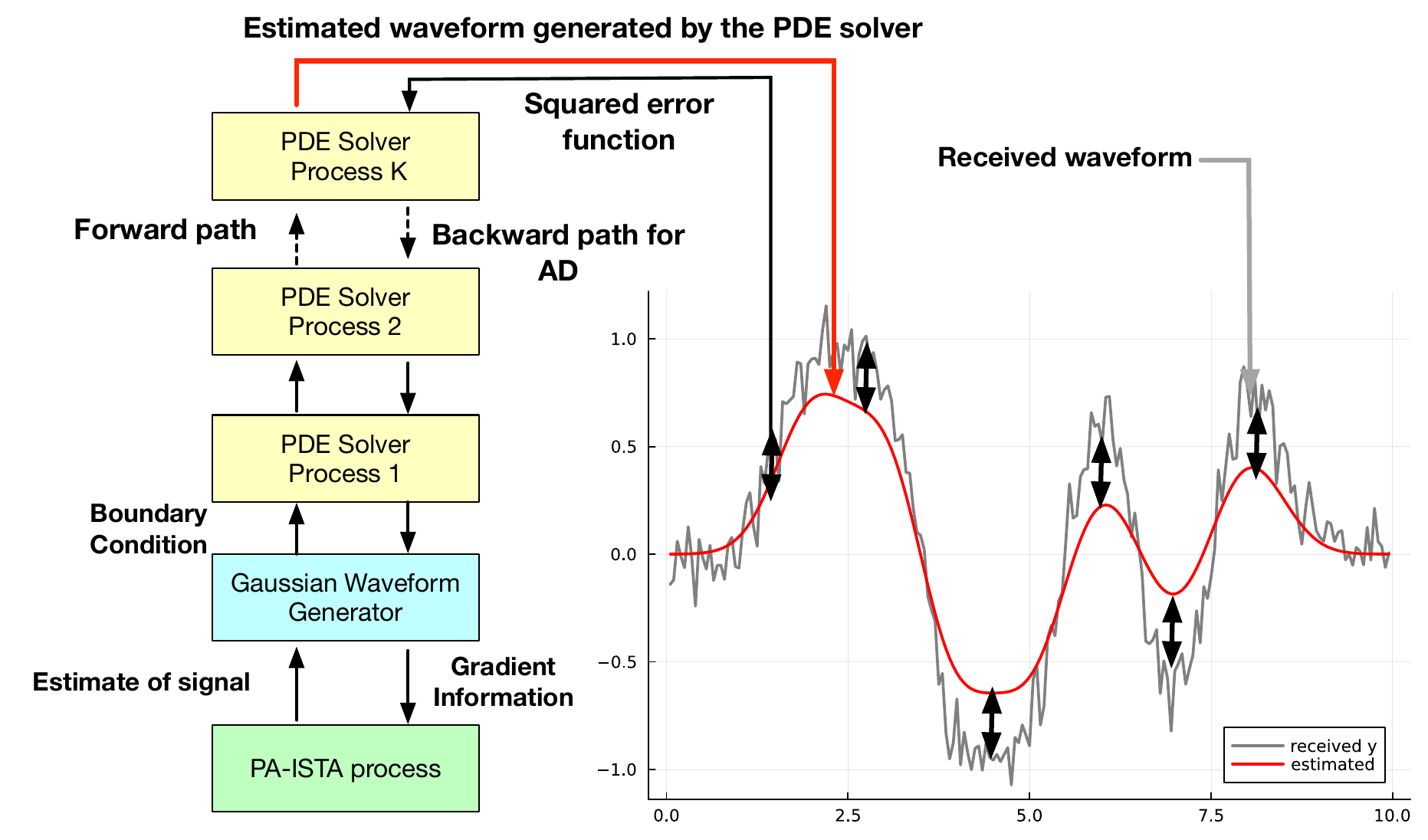}
	\caption{Signal recovery process of PA-ISTA.}
	\label{fig:decoding}
	\end{center}
\end{figure}

The  signal recovery process of PA-ISTA is also depicted in Figure \ref{fig:decoding}. 
From the estimated signal $\bm x^{(k)}$, the initial waveform is generated as a boundary condition of the NLSE.
The SSFM solver then propagates the waveform through the fiber, yielding the predicted output $\hat{f}(\bm x^{(k)})$.
The difference between the predicted output and the observed samples is then used to update the state vector $\bm x^{(k+1)}$.
This process is repeated until the state vector converges to a sparse solution 
or the number of iterations reaches the predefined maximum value.

We initialize the state vector using the zero-forcing solution obtained through
 the {\em digital back propagation(DBP)}
\cite{Ip}, 
a widely established technique for nonlinearity compensation in optical fiber communications. 
DBP reconstructs the input signal by numerically solving the NLSE in reverse, 
with the signs of both dispersion and nonlinearity coefficients inverted. 
This approach effectively "undoing" the channel effects by propagating the received signal backwards 
through a virtual fiber with inverse channel parameters.

In the ideal case of noiseless transmission, DBP can achieve almost perfect signal recovery. 
However, in practical scenarios where noise is present, DBP's performance degrades significantly 
due to its inherent noise amplification characteristics.

 \begin{algorithm}[htbp]
	\caption{Physics-Aware ISTA (PA-ISTA)}
	\label{alg:pa-ista}
	\begin{algorithmic}[1]
	\REQUIRE Observations $\mathbf{y}$, observation position $L$, NLSE parameters $\beta_2$, $\gamma$, step size parameters $\boldsymbol{\eta}$, shrinkage parameters $\boldsymbol{\theta}$, iterations $U$
	\ENSURE Estimated signal $\hat{\mathbf{s}}$
	\STATE $\mathbf{x}^{(0)} \gets \text{DBP}(\mathbf{y}, L, \beta_2, \gamma)$ \COMMENT{Initialize with DBP solution}
	\FOR{$k = 0$ to $U-1$}
		\STATE $b(t) \gets \sum_{i=1}^n x_i^{(k)} \phi(t - p_i)$ \COMMENT{Construct input waveform}
		\STATE Solve the NLSE with boundary condition $$\hat{U}(t, 0;\mathbf{x}^{(k)}) = b(t)$$ using SSFM solver
		\STATE $r_i \gets \hat{U}(q_i, L;\mathbf{x}^{(k)}), \, i \in [m]$ \COMMENT{Generate reconstructed samples}
		\STATE $\mathbf{r}(\mathbf{x}^{(k)}) \gets (r_1, r_2, \ldots, r_m)$ \COMMENT{Construct measurement vector}
		\STATE $\mathbf{g} \gets \nabla_{\mathbf{x}^{(k)}} \|\mathbf{y} - \mathbf{r}(\mathbf{x}^{(k)})\|^2_2$ \COMMENT{Compute Wirtinger gradient by AD}
		\STATE $\mathbf{z}^{(k)} \gets \mathbf{x}^{(k)} - \eta^{(k)}\mathbf{g}$ \COMMENT{Gradient descent step}
		\STATE $\mathbf{x}^{(k+1)} \gets \mathcal{T}_{\theta^{(k)}}(\mathbf{z}^{(k)})$ \COMMENT{Shrinkage step}
	\ENDFOR
	\STATE $\hat{\mathbf{s}} \gets \mathbf{x}^{(U)}$ \COMMENT{Return final estimate}
	\RETURN $\hat{\mathbf{s}}$
	\end{algorithmic}
	\end{algorithm}

 \subsection{Deep Unfolding (DU) for Parameter Tuning}
 
The performance of PA-ISTA critically depends on the choice of two key parameters: the step size parameter $\eta^{(k)}$ and 
the shrinkage parameter $\theta^{(k)}$.
These parameters significantly influence both the convergence behavior and the quality of the estimated signal. Unlike conventional ISTA for linear measurements, where theoretical guidelines exist for parameter selection (e.g., $\eta	< 1/\Gamma$), the nonlinear nature of our physics-aware framework makes it challenging to establish general rules for parameter setting.
To address this challenge, we propose to leverage {\em deep unfolding (DU)}\cite{LISTA, 9020494, Borgerding}, 
a technique that bridges iterative optimization algorithms and deep learning. 
Deep unfolding treats each iteration of an optimization algorithm as a layer in a neural network, 
allowing the algorithm's parameters to be learned from training data. 
In this approach, the iterations of PA-ISTA are ``unfolded'' into a fixed-depth network structure, 
where each layer maintains the same form as a PA-ISTA iteration but with learnable parameters. 
This transformation enables us to optimize the step size and shrinkage parameters 
through standard neural network training procedures while preserving the physics-aware nature of the algorithm.

\subsubsection{Nested structure of gradient computation}
Applying deep unfolding to PA-ISTA presents a unique technical challenge due to the {\em nested structure of gradient computations}. In each iteration of PA-ISTA, AD is used to compute gradients through the physics model with the SSFM solver. When we attempt to unfold PA-ISTA into a neural network structure, we encounter a situation where AD needs to be performed at two different levels: one for the physics model within each iteration, and another for the end-to-end training of the unfolded network.
This nested AD structure poses a significant implementation challenge 
because most of AD do not support nested AD computations. 
i.e., typical AD engines are designed to handle a single level of gradient computation.
This limitation creates a fundamental obstacle for parameter tuning through DU, as we cannot directly apply standard neural network training procedures to our unfolded PA-ISTA architecture. Resolving this nested AD challenge is crucial for successfully implementing the deep unfolding approach for PA-ISTA parameter optimization.

\begin{figure*}[htbp]
	\begin{center}
	\includegraphics[width=1.5\columnwidth]{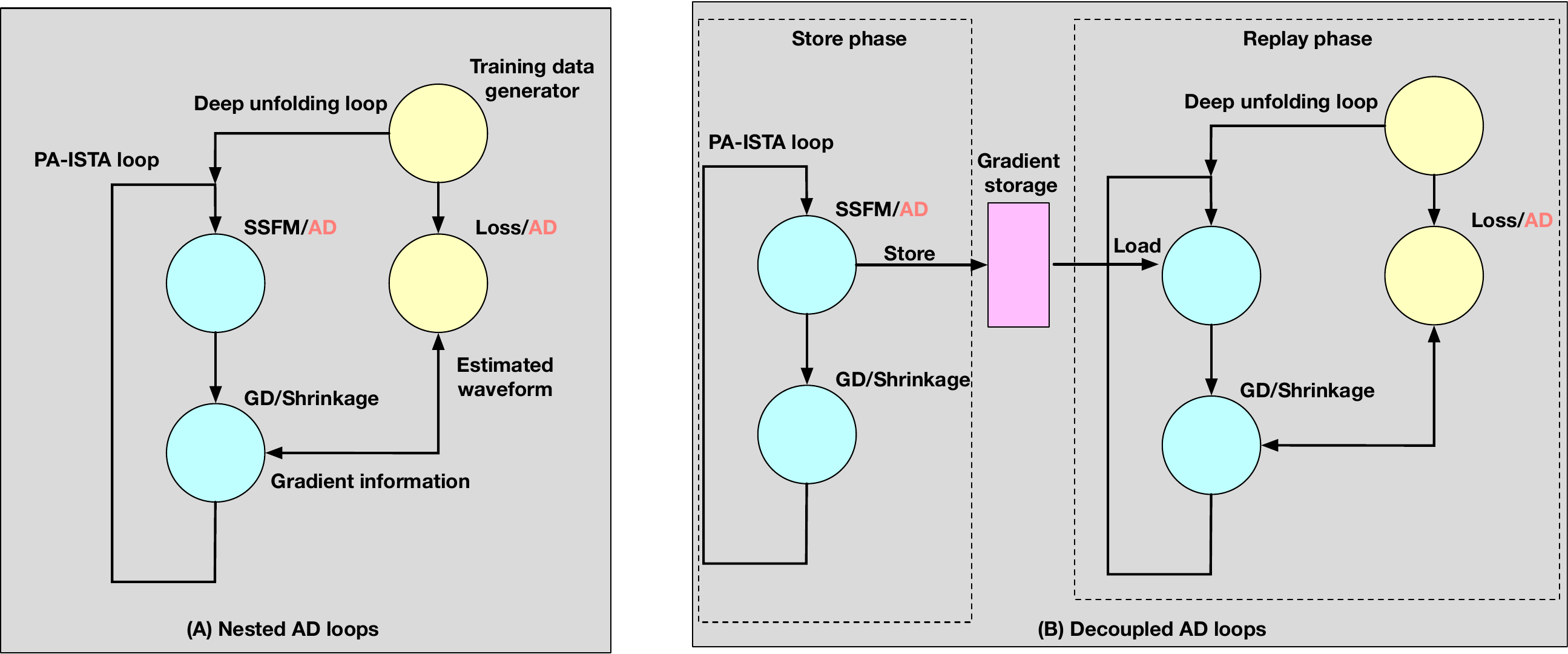}
	\caption{Nested structure of gradient computation (A) and store-and-replay method (B).}
	\label{fig:nested_ad}
	\end{center}
\end{figure*}

\subsubsection{Store-and-Replay Method}
To overcome this implementation challenge, we propose the {\em store-and-replay method}, which effectively decouples the two levels of AD by executing PA-ISTA iterations twice in a specific manner. The key idea is to separate the physics-based Wirtinger gradient computation from the DU-parameter optimization process.
Algorithm \ref{store_and_replay} presents a pseudo code of the store-and-replay method. 
In the first pass (Store phase), 
we execute PA-ISTA iterations with AD enabled for the Wirtinger derivative.
The second pass (Replay phase) then replicates the same PA-ISTA iterations 
but with a crucial difference; the gradients are  retrieved from the storage.
This two-pass approach successfully avoids nested AD while maintaining the mathematical equivalence 
to the original algorithm. The store-and-replay method enables us to apply DU for parameter tuning 
by effectively separating the physics-based gradient computation from the DU parameter optimization process. 
The details of the DU-training process for PA-ISTA is given in Algorithm \ref{alg:du-pa-ista}.

\begin{algorithm}[htbp]
 \caption{Pseudo code of store-and-replay method}
 \label{store_and_replay}
 \begin{algorithmic}[1]
  \STATE Iterate the following two phases with a distinct pair $(\bm y, \bm s)$.
  \STATE {\bf [Store phase]} Compute Wirtinger gradients through the SSFM by AD.
  \STATE Store these gradient vectors in a storage.
  \STATE Execute standard PA-ISTA updates (without DU-training).
  \STATE {\bf [Replay phase]} Retrieve the pre-computed gradient vectors from the storage. Note that the initial state vector  must be exactly the same as that used in the store phase.
  \STATE Use these stored gradients for the PA-ISTA updates.
  \STATE Enable DU parameter learning during replay phase.
 \end{algorithmic} 
 \end{algorithm}

 \begin{algorithm}
	\caption{Training procedure for PA-ISTA with deep unfolding}
	\label{alg:du-pa-ista}
	\begin{algorithmic}[1]
	\REQUIRE Target signal $\mathbf{s}_0$, observations $\mathbf{y}$, observation position $z_{obs}$, learnable parameters $\boldsymbol{\eta}$, $\boldsymbol{\theta}$, optimizer $opt$, iterations $U$, NLSE parameters $\beta_2$, $\gamma$
	\ENSURE Updated parameters $\boldsymbol{\eta}$, $\boldsymbol{\theta}$
	\STATE $\mathbf{s} \gets \text{DBP}(\mathbf{y}, z_{obs}, \beta_2, \gamma)$ \COMMENT{Initialize with BP solution}
	
	\STATE // Store Phase - Compute and store gradients
	\STATE $\nabla\text{History} \gets \text{zeros}(n, U)$ \COMMENT{Gradient storage}
	\FOR{$k = 1$ to $U$}
		\STATE $f(\mathbf{x}) = \|\mathbf{y} - \text{SSFM}(\mathbf{x}, z_{obs}, \beta_2, \gamma)\|^2_2$ \COMMENT{Define loss function}
		\STATE $\nabla\mathbf{s} \gets \text{WirtingerGrad}(f, \mathbf{s})$ \COMMENT{Compute gradient via AD}
		\STATE $\nabla\text{History}[:,k] \gets \nabla\mathbf{s}$ \COMMENT{Store gradient}
		\STATE $\mathbf{s} \gets \mathbf{s} - |\eta_k| \cdot \nabla\mathbf{s}$ \COMMENT{Gradient step}
		\STATE $\mathbf{s} \gets \mathcal{T}_{\theta_k}(\mathbf{s})$ \COMMENT{Shrinkage step}
	\ENDFOR
	
	\STATE // Replay Phase - Update parameters using stored gradients
	\STATE $\mathbf{s} \gets \text{DBP}(\mathbf{y}, z_{obs}, \beta_2, \gamma)$ \COMMENT{Reset initial state}
	\STATE $\mathcal{L}(\boldsymbol{\eta}, \boldsymbol{\theta}) = \|\mathbf{s}_U - \mathbf{s}_0\|^2_2$ \COMMENT{Define parameter loss}
	\STATE Compute gradients $\nabla_{\boldsymbol{\eta}}\mathcal{L}$, $\nabla_{\boldsymbol{\theta}}\mathcal{L}$ via backpropagation through:
	\FOR{$k = 1$ to $U$}
		\STATE $\mathbf{s} \gets \mathbf{s} - |\eta_k| \cdot \nabla\text{History}[:,k]$ \COMMENT{Use stored gradients}
		\STATE $\mathbf{s} \gets \mathcal{T}_{\theta_k}(\mathbf{s})$ \COMMENT{Apply shrinkage}
	\ENDFOR
	\STATE Update $\boldsymbol{\eta}$, $\boldsymbol{\theta}$ using an optimizer (e.g., Adam) with gradients $\nabla_{\boldsymbol{\eta}}\mathcal{L}$, $\nabla_{\boldsymbol{\theta}}\mathcal{L}$
	\STATE \RETURN Updated parameters $\boldsymbol{\eta}$, $\boldsymbol{\theta}$
	\end{algorithmic}
	\end{algorithm}
	
	\subsection{Convergence and Computational Complexity}
	\label{subsec:conv_complex}
	
	\subsubsection{Convergence}
	Consider the objective
$$
	F(\bm s)=\tfrac{1}{2}\|\bm y-\hat f(\bm s)\|_2^2+\lambda\|\bm s\|_1,
$$
	where \(\hat f\) is the differentiable forward map obtained by the SSFM-based solver and AD (Wirtinger calculus). 
	Although \(F\) can be nonconvex due to the nonlinear forward operator, the data-fidelity gradient \(\nabla d(\bm s) \equiv\nabla\tfrac{1}{2}\|\bm y-\hat f(\bm s)\|_2^2\) is locally Lipschitz under a mild regularity assumption on \(\hat f\) (bounded Jacobian on the level set of interest). 
	Let \(\{\bm s^{k}\}\) be generated by the PA-ISTA iteration
	\begin{align}
	\bm s^{k+1}=\mathrm{prox}_{\eta_k\lambda\|\cdot\|_1}\!\big(\bm s^k-\eta_k\nabla d(\bm s^k)\big)
	\end{align}
	with either (i) backtracking to ensure a sufficient decrease condition, or (ii) fixed steps \(\eta_k\in(0,\bar\eta]\) with \(\bar\eta<1/L\) where \(L\) is a local Lipschitz constant of \(\nabla d\).
	For backtracking, we can shrink $\eta_k\leftarrow\tau\,\eta_k$ with $\tau\in(0,1)$ until $F(\bm s^{k+1}) \le F(\bm s^{k}) - \frac{\sigma}{2\eta_k}\,\|\bm s^{k+1}-\bm s^{k}\|_2^2$ holds for some fixed $\sigma\in(0,1)$.
	Then every limit point of \(\{\bm s^k\}\) is a \emph{first-order critical point} of \(F\) (proximal-stationary point), and \(F(\bm s^{k})\) is nonincreasing. 
	This is the standard proximal-gradient convergence behavior for smooth\,+\,nonsmooth composite objectives and remains valid in the present setting because the nonsmooth part is convex (\(\ell_1\)) and the gradient of the smooth part is (locally) Lipschitz.
	Deep unfolding keeps the iteration form but removes a priori guarantees unless the learned stepsizes obey a safeguard (e.g., $\eta_k\le (1-\epsilon)/L$). With either such a safeguard or backtracking ensuring sufficient decrease, $F(\bm s^k)$ is nonincreasing and every limit point is proximal-stationary.

	\subsubsection{Computational complexity}
	The proposed algorithm exhibits a double-loop structure: an outer loop implementing the proximal gradient iterations, 
	and an inner loop executing the SSFM. The computational complexity of the SSFM solver scales with the number of grid points, 
	requiring $O(N_t N_z)$ operations, where $N_t$ and $N_z$ represent the number of grids in $t$ and $z$ direction, respectively.
   The overall computational complexity of the algorithm becomes $O(U N_t N_z \log N_t)$ where
   $N_t \log N_t$ comes from the complexity of FFT.
	A practical trade-off exists between computational efficiency and solution accuracy. 
	While using a coarser grid (smaller $N_t$ and $N_z$) reduces computational time, 
	it may compromise the quality of the recovered signal. Therefore, careful selection of grid parameters is crucial 
	for balancing computational complexity against recovery performance.

	The memory use is dominated by the SSFM state and AD tapes. 
	The \emph{store-and-replay} strategy decouples nested AD, so peak memory is \(\mathcal{O}(N_t)\) per SSFM slice plus \(\mathcal{O}(nU)\) for the stored Wirtinger gradients, while runtime remains linear in \(U,N_t,N_z\). 
	Therefore, PA-ISTA exhibits near-linear scaling in grid resolution and iteration count, and its wall-clock can be traded off against accuracy through the choice of \((N_t,N_z)\) and \(U\), as documented in the method description.

\subsection{QPSK detection with PA-ISTA}
The shrinkage function is a crucial component that incorporates prior information about the source signal, 
such as sparsity, into the recovery process. 
Although soft thresholding is commonly used for sparse signal recovery, 
one of PA-ISTA's key strengths is its ability to handle diverse source signals 
through appropriate modifications to the shrinkage function. 
By simply adjusting this function within the PA-ISTA framework, 
we can effectively leverage different types of prior information about the characteristics of the source signal.

We here consider an application of PA-ISTA for quadrature phase-shift keying (QPSK) detection. In this scenario, the input signal 
is assumed to be a QPSK signal, and the observation is the received signal at the detector. 
The goal is to estimate the transmitted signal from the observation. This is a simpliefied 
scenario of the QPSK detection problem, and the PA-ISTA can be applied to this problem.
The prior knowledge of the QPSK signal can be reflected by the shrinkage function. 
We can expect that the shrinkage function is matched to the QPSK signal provides better detection performance
compared with the case of a naive DBP-based recovery.

Assume that the QPSK signal constellation is given by
\begin{align}
\mathcal{Q} \equiv \left\{ 1+i, -1+i, -1-i, 1-i \right\},
\end{align}
i.e., a transmitted signal is chosen from the set $\mathcal{Q}$ with equal probability.
The initial waveform is generated from the QPSK signal with equal probability:
\begin{align}
	U(\tau, 0) =  \sum_{i=1}^n s_i \phi(\tau - p_i),
\end{align}
where $s_i \in \mathcal{Q}$ and $p_i$ is the position of the $i$-th element.
The rest of the observation process is the same as the case of the sparse signal recovery discussed in the previous section.

In this case, the process of the PA-ISTA is modified as
\begin{align}
\bm{z}^{(k)} &= \bm{x}^{(k)} - \eta^{(k)} \nabla_{\bm{x}^{(k)}} \|\bm y - \hat{f}(\bm{x}^{(k)}) \|^2_2 \\
\bm{x}^{(k+1)} &= S(\bm{z}^{(k)}, \theta^{(k)}), \quad k=0,1,\dots, 
\end{align}
where $S:\mathbb{C} \times \mathbb{R} \rightarrow \mathbb{C}$ is the shrinkage function matched to the QPSK signal. 
The shrinkage function is given by
\begin{align}
	S(y,\lambda) &\equiv \tanh(\lambda \Re(y)) + i\tanh(\lambda \Im(y)),
\end{align}
where $\Re(y)$ and $\Im(y)$ denote the real and imaginary parts of $y$, respectively.
The parameter $\lambda$ is a scaling factor that controls the strength of the shrinkage.
The final estimate is given by  
\begin{align}
\hat{\bm s} = \operatorname{proj}_{\mathcal{Q}}(\bm x^{(U)}),
\end{align}
where 
$\operatorname{proj}_{\mathcal{Q}}(a+ib)=\operatorname{sign}(a)+i\,\operatorname{sign}(b)$.

The above detection process can be regarded as a projected gradient descent method with 
the soft projection operator given by the set $\mathcal{Q}$. In the context of MIMO signal detection,
such a detection process is discussed in \cite{Takabe}. 
In optical fiber communications, signal detection is a key process for the receiver design.
An iterative algorithm like PA-ISTA can be used for the signal detection to improve the detection performance.

\section{Numerical experiments} 
\label{sec:experiments}

\subsection{Signal Recovery Performance of PA-ISTA}

We conducted extensive numerical experiments to evaluate the signal recovery performance of PA-ISTA. The simulations were performed using the following parameter setting.
For the NLSE parameters, we set the dispersion constant $\beta_2$ to $-10$ and
the nonlinear coefficient $\gamma$ to $2$. The pulse half-width $T_0$ was set to 1, resulting in a dispersion length $L_D \equiv T_0^2/|\beta_2| = 0.1$, and nonlinear length $L_{NL} \equiv 1/\gamma = 0.5$.
These parameters were chosen to represent typical propagation conditions where both dispersion and nonlinear effects significantly influence the signal evolution.
The SSFM solver was configured with a spatial step size ($z$-direction) of $\Delta z = 0.01$ and a temporal step size 
($t$-direction) of $\Delta t = 0.3$. The temporal grid consisted of $N_t=256$ points spanning the interval $[-38.4, 38.4]$, providing sufficient resolution for accurate waveform propagation simulation.
For our experiments, we generated sparse test vectors with the following parameters. Each test vector had length $n = 30$ with exactly $k = 3$ non-zero elements, representing a sparse signal. The positions of the non-zero elements were randomly distributed across the vector following a uniform distribution. The values of the non-zero elements were complex numbers, each with unit magnitude ($|s_i| = 1$) but with uniformly randomly chosen phases. 

The signal propagation and measurement were simulated over a distance of $L = 3L_D = 0.3$. To generate received signals, we synthesized the received waveform using the SSFM solver with the previously specified parameters. This approach allows us to evaluate the recovery performance under well-controlled conditions while maintaining the essential nonlinear characteristics of optical fiber propagation. We used 	$q_i = -38.4 + ih, i = 0,1,\ldots,255$ as sensing positions. The simulation code is implemented using Julia 1.9 and AD mechanism in Zygote.jl.

The deep unfolding training was conducted under the following experimental conditions. 
We define the complex AWGN as $n_i\sim\mathcal{CN}(0,\sigma^2)$ and the SNR by
\[
\mathrm{SNR\,[dB]}\equiv 10\log_{10}\!\left(
\frac{1}{\sigma^2}
\right).
\]
In the following experiments, we set SNR to $15$ dB. The network was trained for 100 iterations, with PA-ISTA configured to perform 30 iterations per forward pass. The step size parameters $\eta^{(k)}$ were initialized to $0.01$, and the shrinkage parameters $\theta^{(k)}$ were initialized to $0.001$ for $k=0,1,\ldots,29$. These initial parameters were found by an ad hoc manner.
For optimization of the learnable  parameters, we employed the Adam optimizer \cite{Adam} with a learning rate of $10^{-4}$ to simultaneously update both  $\eta^{(k)}$ and 
	$\theta^{(k)}$. The deep unfolding architecture was implemented following the methodology described in \cite{TISTA}.

\begin{figure}[htbp]
\begin{center}
\includegraphics[width=0.9\columnwidth]{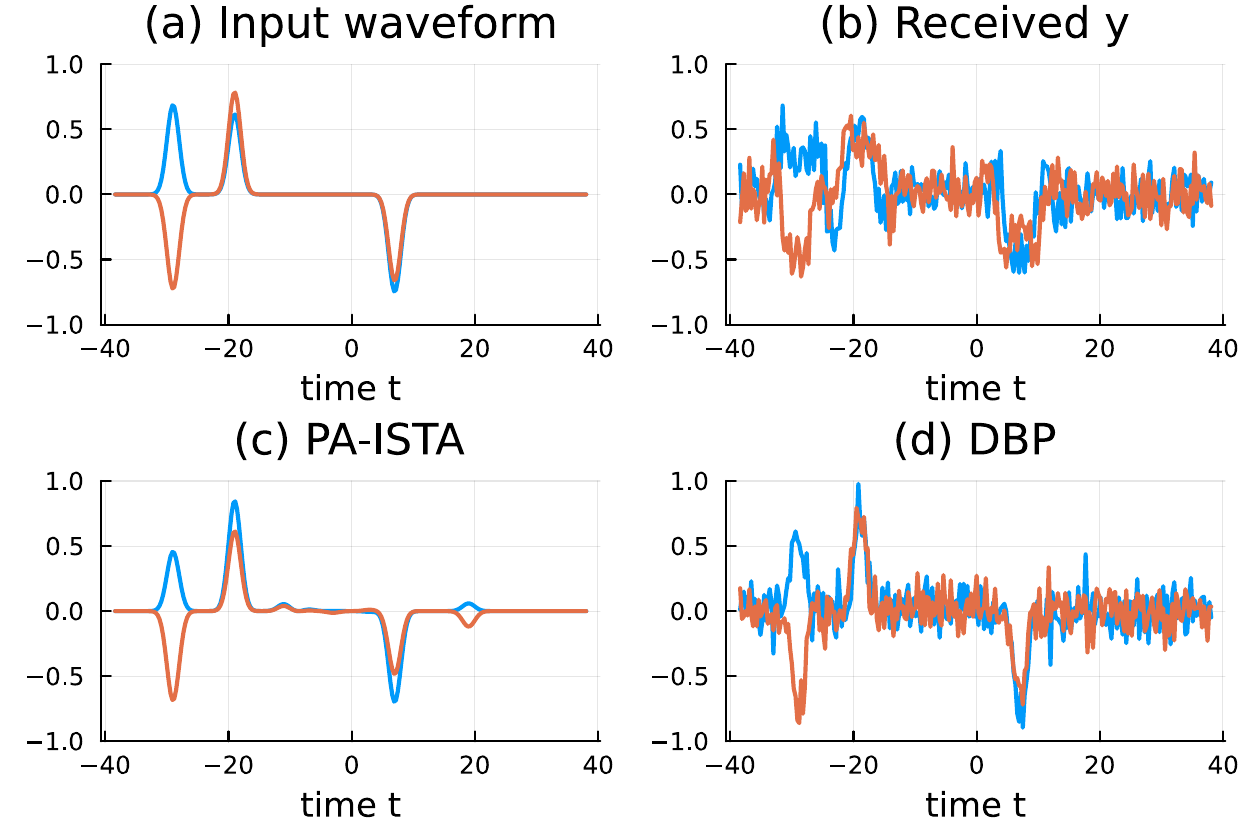}
\caption{Example of signal recovery. PA-ISTA with DU-optimized parameters were used (Blue:real part, Red: imaginary part). The SNR was set to 15 dB.}
\label{fig:wave}
\end{center}
\end{figure}

Figure \ref{fig:wave} presents an example of sparse signal recovery using PA-ISTA with parameters optimized through deep unfolding. 
Panel (c) shows the recovered signal, which closely matches the original input waveform shown in panel (a), demonstrating the effectiveness of our approach. For comparison, panel (d) shows the result of conventional DBP-based recovery. The comparison clearly illustrates that DBP fails to provide robust recovery in the presence of observation noise, whereas PA-ISTA maintains reliable performance under these challenging conditions.

\begin{figure}[htbp]
\begin{center}
\includegraphics[width=0.9\columnwidth]{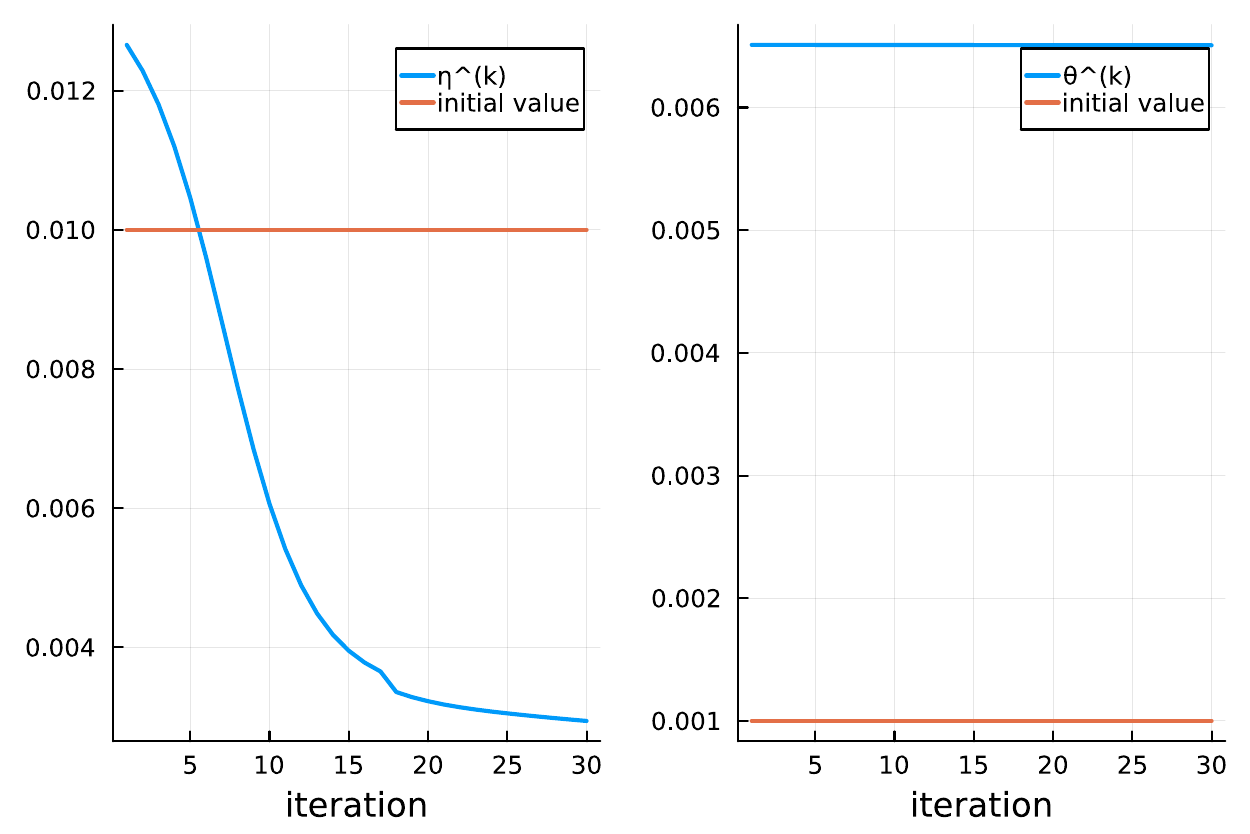}
\caption{Tuned parameters by deep unfolding.}
\label{fig:params}
\end{center}
\end{figure}

Figure \ref{fig:params} illustrates the values of the optimized parameters $\eta^{(k)}$ and $\theta^{(k)}$, 
alongside their initial values before training. 
A notable observation is the time-varying behavior of the optimized step size $\eta^{(k)}$, 
which suggests that different step sizes are beneficial at different stages of the recovery process. 
This adaptive nature of the parameters, discovered through DU, contrasts with conventional approaches 
that typically employ fixed parameters throughout the iterations.

\begin{figure}[htbp]
\begin{center}
\includegraphics[width=0.9\columnwidth]{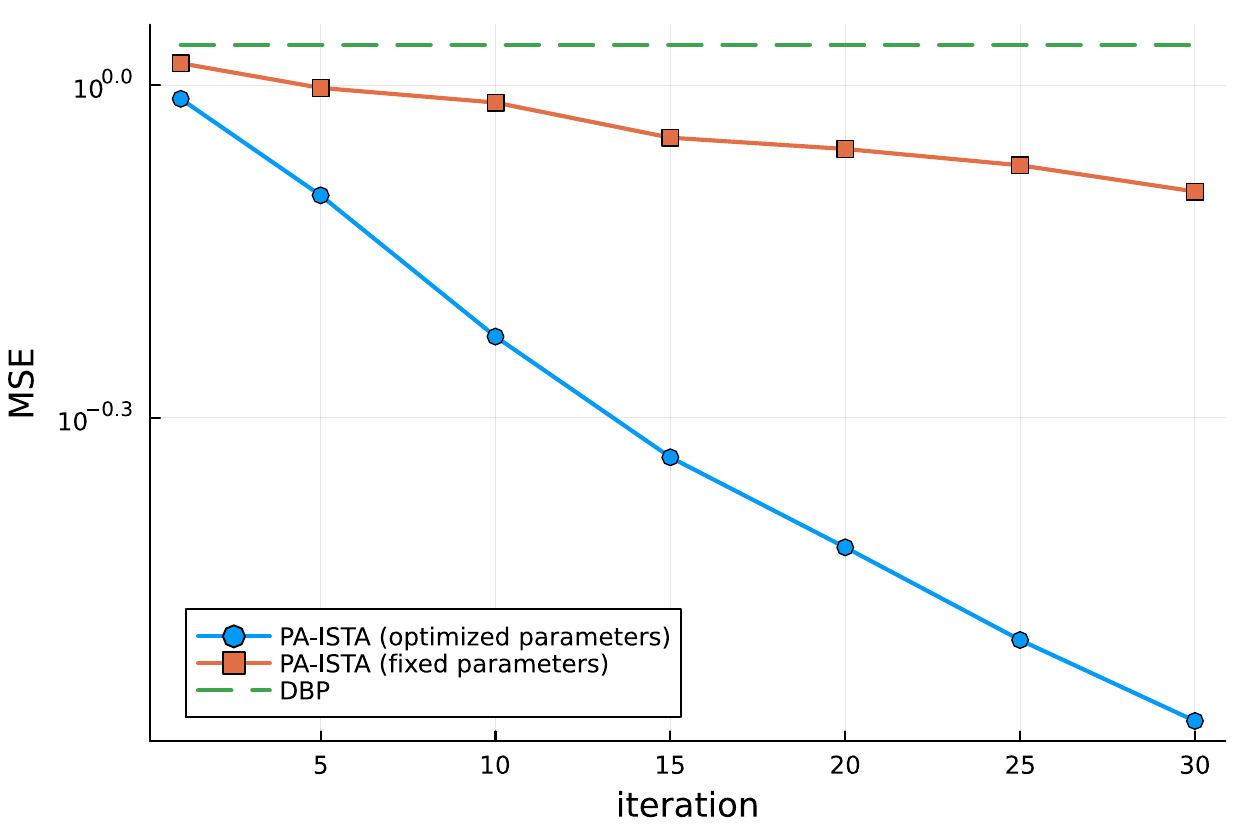}
\caption{Comparisons of MSE performance of PA-ISTA. }
\label{fig:MSE}
\end{center}
\end{figure}

Figure \ref{fig:MSE} compares the mean squared error (MSE) performance of PA-ISTA under different SNR conditions (5 dB). The MSE is defined as ${\sf E}\left[\|\bm s - \hat{\bm s}\|_2^2 \right]$, where $\bm s$ represents the true sparse vector and $\hat{\bm s}$ is the PA-ISTA estimate.
The expectation is estimated from 100 independent trials.
The results demonstrate that PA-ISTA achieves substantially lower MSE values compared to conventional DBP-based methods, whose performance levels are indicated by horizontal lines in the figure. This significant performance gap clearly highlights the advantages of our physics-aware approach over conventional recovery methods.
The figure also includes MSE curves for PA-ISTA using fixed initial parameters without DU optimization. These curves exhibit notably slower convergence, as evidenced by their shallower slopes compared to the versions with optimized parameters. This comparison provides compelling evidence for the effectiveness of DU in enhancing reconstruction performance. The optimization of iteration-dependent parameters through DU leads to both faster convergence and better final recovery accuracy.

\subsection{QPSK Detection with PA-ISTA}

In this subsection, we present the numerical results of the QPSK detection problem with PA-ISTA.
The basic setup of the QPSK detection problem follows the same configuration 
as in the sparse signal recovery case.
The parameters of the NLSE model were set to $\beta_2 = -10$ and $\gamma = 2$.
The pulse half-width $T_0$ was set to 1, resulting in a dispersion length 
$L_D \equiv T_0^2/|\beta_2| = 0.1$, and nonlinear length $L_{NL} \equiv 1/\gamma = 0.5$.
The SSFM solver was configured with a spatial step size ($z$-direction) of $\Delta z = 0.01$ and 
a temporal step size ($t$-direction) of $\Delta t = 0.3$. The temporal grid consisted of $N_t =256$ points spanning 
the interval $[-38.4, 38.4]$, providing sufficient resolution for accurate waveform propagation simulation. The signal propagation and measurement were simulated over a 
distance of $L = 5L_D = 0.5$.
The length of a QPSK signal vector is $n = 15$. 
Deep unfolding was used to optimize the parameters of PA-ISTA.

\begin{figure}[htbp]
\begin{center}
\includegraphics[width=0.9\columnwidth]{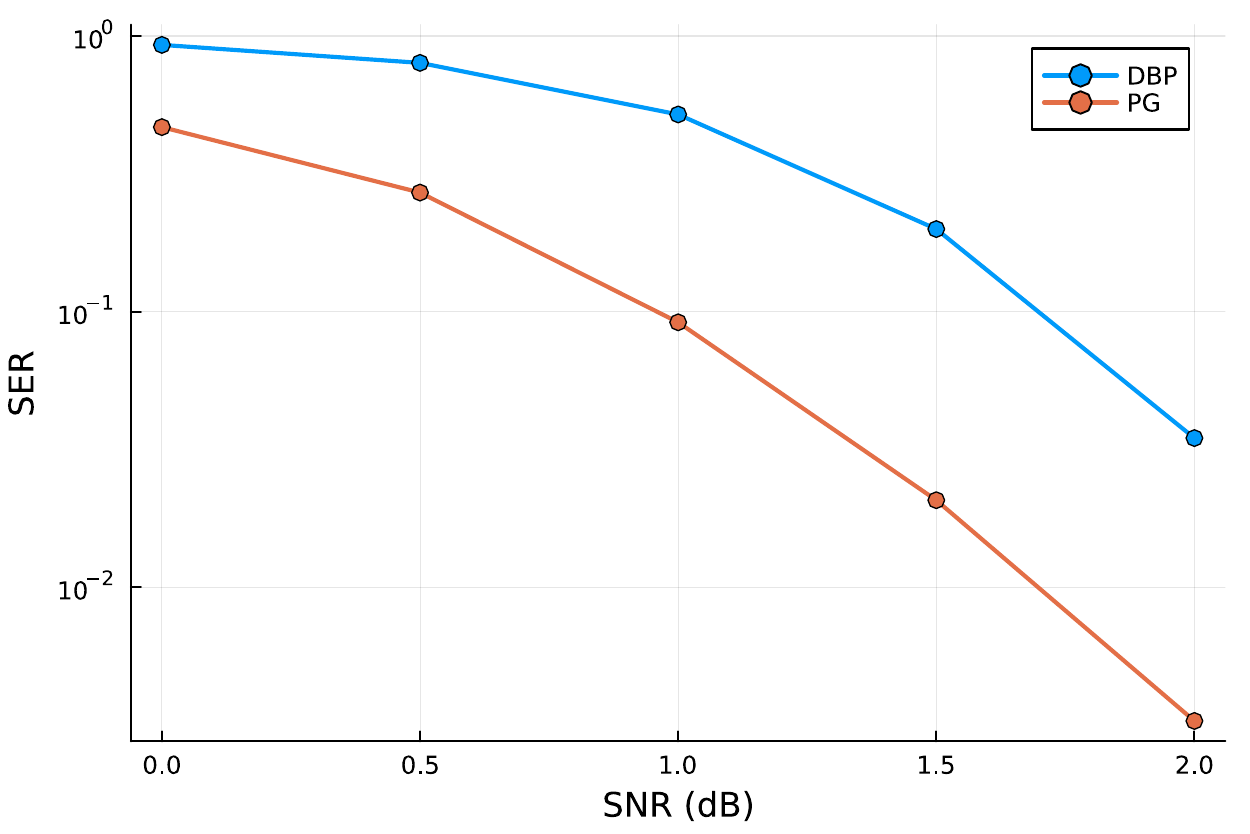}
\caption{Symbol error rate (SER) performance of PA-ISTA for QPSK detection.}
\label{fig:QPSK_PA-ISTA}
\end{center}
\end{figure}

The symbol error rate (SER) performance of PA-ISTA is shown in Figure \ref{fig:QPSK_PA-ISTA}.
As a benchmark, we also show the SER performance of backpropagation-based recovery $DBP(\bm y)$.
We can observe that PA-ISTA achieves much better SER performance than $DBP(\bm y)$ in all SNR conditions.
The performance gap between PA-ISTA and $DBP(\bm y)$ can be explained by the use of prior knowledge 
of the QPSK signal in the shrinkage function.
The results indicate that the PA-ISTA can be a promising approach for the QPSK detection problem.

\section{Conclusion} 
\label{sec:conclusion}

This paper has presented a physics-aware framework for sparse signal recovery with PDE-governed measurement systems. 
By incorporating physical models directly into the recovery process through AD and numerical PDE solvers, 
our approach demonstrates the potential of physics-aware signal processing. 
The core idea is also shown to be effecitive in decoding process for binary linear codes in PDE-governed channels as well \cite{ISIT2025}.
The key innovation lies in the seamless integration of physical models with a gradient descent process through AD, 
enabling the algorithm to utilize gradient information from the PDE solver effectively.
The proposed PA-ISTA algorithm, while illustrated through optical fiber applications,
establishes a general methodology for integrating physical constraints into 
sparse recovery problems with PDE-governed measurement systems.

Our numerical experiments with optical fiber channels demonstrate that PA-ISTA significantly outperforms 
conventional recovery methods in terms of MSE in sparse signal recovery, 
and symbol error rate performance in QPSK detection.
These results validate the practical viability of our physics-aware approach for real-world applications.

The primary future challenge lies in the computational complexity introduced by the double-loop structure of 
our method. Although using coarser grids can partially address this issue, 
balancing computational efficiency with solution accuracy remains a crucial challenge for practical implementations.
Additionally, extending our framework to handle time-varying physical systems and developing 
theoretical performance guarantees remain important open problems.
As computational capabilities continue to advance, 
physics-aware approaches may become increasingly practical for a wider range of applications in sensing and communication systems.

\section*{Acknowledgment}
This work is supported by JST CRONOS (JPMJCS25N5).

\end{document}